\documentstyle[12pt]{article}

\newcommand{\T}{T_{\mbox{\scriptsize vac}}^{\mu\nu}}
\renewcommand{\O}{{\cal O}}
\renewcommand{\u}{u=\mbox{const.}}
\renewcommand{\v}{v=\mbox{const.}}
\newcommand{\us}{u=\mbox{\scriptsize const.}}
\newcommand{\Cv}{v_{\mbox{\scriptsize crit}}}
\newcommand{\rr}{(\nabla r)^2}
\newcommand{\rv}{(\nabla r,\nabla v)}
\newcommand{\ru}{(\nabla r,\nabla u)}
\newcommand{\uv}{(\nabla u,\nabla v)}
\newcommand{\dr}{\frac{\partial r(u,v)}{\partial u}}
\newcommand{\AHplus}{\mbox{AH}^+}
\newcommand{\AHmin}{\mbox{AH}^-}
\newcommand{\AHpm}{\mbox{AH}^\pm}
\newcommand{\rHplus}{r^+_{\mbox{\tiny AH}}}
\newcommand{\rHpm}{r_{\mbox{\tiny AH}}^{\pm}}
\newcommand{\BHplus}{B_{\mbox{\tiny AH}}^+}
\newcommand{\I}{{\cal I}^+}
\newcommand{\ein}{\varepsilon_{\mbox{\scriptsize in}}}
\newcommand{\eout}{\varepsilon_{\mbox{\scriptsize out}}}
\newcommand{\wkappa}{\widetilde{\kappa}}
\newcommand{\wk}{\widetilde{k}}
\begin{document}

\begin{center}
{\LARGE\bf
Post-radiation evolution of black holes}
\end{center}
\begin{center}
{\bf G.A. Vilkovisky}
\end{center}
\begin{center}
Lebedev Physical Institute, and Lebedev
Research Center in Physics,\\ Leninsky Prospect 53, 119991 Moscow,
Russia.\\ E-mail: vilkov@sci.lebedev.ru \\
\end{center}
\vspace{3cm}
\begin{abstract}

The expectation-value equations for the collapse of a
macroscopic, spherically symmetric, and uncharged body
are integrated up to the limit of validity of 
semiclassical theory. The collapse finishes with a true
stable black hole of the mass microscopically exceeding
the vacuum-induced charge. The apparent horizon is almost
closed. The most important feature of the solution is the
presence of an irremovable Cauchy horizon.
\end{abstract}
\newpage
$$            $$

The present paper is a sequel of Ref. [1] in which the
expectation-value problem is considered for the 
gravitational collapse of a spherically symmetric
uncharged source having a compact spatial support and
a macroscopic ADM mass $M$. The evolution equations for
the metric have been integrated in Ref. [1] only up to
the instant of retarded time at which the Hawking
radiation terminates, but the problem of the further
integration was not that the limit of validity of
semiclassical theory was reached. Rather, some of the
techniques used in Ref. [1] ceased working. Here these
difficulties are surmounted, and the solution is
pushed to the limit of validity of semiclassical theory.

The notation in the present paper is the same as in 
Ref. [1] except for $\lambda$ in Eq. (25) below.
Specifically, $\v$ and $\u$ are the equations of the
radial past and future light cones, and
\begin{equation}
A=4\pi r^2\left(\T \nabla_\mu u\nabla_\nu u\right)
\frac{\rv^2}{\uv^2}\;,
\end{equation}
\begin{equation}
D=4\pi r^2\left(\T \nabla_\mu v\nabla_\nu v\right)
\frac{1}{\rv^2}\;,
\end{equation}
\begin{equation}
T_1=4\pi r^2\left(\T \nabla_\mu u\nabla_\nu v\right)
\frac{2}{\uv}\;,
\end{equation}
\begin{equation}
T_1+T_2=4\pi r^2\T\, g_{\mu\nu}\;,
\end{equation}
\begin{equation}
B=1-\rr+T_1=r\triangle r\;.
\end{equation}
Here $\T$ is the energy-momentum tensor of the in-vacuum,
and $\triangle$ is the D'Alembert operator in the
Lorentzian subspace of a spherically symmetric spacetime.
Below, both $v$ and $u$ are normalized so as to be the
proper time of an observer at infinity. $\O$ is the notation
for a quantity that vanishes as $\mu/M\to 0$, and $\mu$ is
the Planckian mass ($1$ in the absolute units). An inequality
of the form $X>|\O|$ assumes any $\O$ and signifies that
$X$ is a macroscopic quantity.

As discussed in Ref. [1], in the semiclassical region [1]
the expectation value of the metric satisfies the equations
\begin{equation}
T_1+T_2=\O\;,\qquad A=\O\;.
\end{equation}
It has been shown in Ref. [1] that, as far as the solution
obtained in this reference extends, it satisfies also
the equations
\begin{equation}
T_1+T_2=\O\;,\qquad D=\O\;.
\end{equation}
A necessary condition for the equivalence of Eqs. (6) and (7)
is that the data functions be slowly varying with respect
to the time of an observer at infinity. No signs of a
violation of this condition are observed in the solution.
The obstacle encountered in Ref. [1] was the appearance
in the solution of a line $B=0$. The appearance of this
line at sufficiently late $u$ is directly related to the
vanishing of the radiation temperature. The idea of the
present work is to use Eqs. (7) to get across $B=0$.

Both sets of equations, (6) and (7), are valid globally but
neither of them can be used to obtain the solution globally.
The solution can be obtained in two complementary regions.
Eqs. (6) can be solved in the region
\begin{equation}
\rr>O\left(\sqrt{|A|}\right)
\end{equation}
called in [1] the region of weak field, and the solution is [1]
\begin{equation}
\rr=1-\frac{2{\cal M}(u)}{r}+\frac{Q^2(u)}{r^2}\;,
\end{equation}
\begin{equation}
\ru=-1+\O\;,
\end{equation}
\begin{equation}
B=\frac{2}{r}\left({\cal M}(u)-\frac{Q^2(u)}{r}\right)\;,
\end{equation}
\begin{equation}
D=\frac{4}{\left(\rr\right)^2}
\left(-\frac{d{\cal M}(u)}{du}+
\frac{1}{2r}\frac{dQ^2(u)}{du}\right)
\end{equation}
where ${\cal M}(u)$ and $Q^2(u)$ are arbitrary data functions.

Eqs. (7) can be solved in the region
\begin{equation}
r<\frac{1}{|\O|}
\end{equation}
which covers the region of strong field
\begin{equation}
\left(\rr<0\right)\bigcup\left(\rr=|\O|\right)\;.
\end{equation}
The solution is
\begin{equation}
\rr=1-\frac{2m(v)}{r}+\frac{q^2(v)}{r^2}\;,
\end{equation}
\begin{equation}
B=\frac{2}{r}\left(m(v)-\frac{q^2(v)}{r}\right)\;,
\end{equation}
\begin{equation}
A=\frac{dm(v)}{dv}-
\frac{1}{2r}\frac{dq^2(v)}{dv}\;,
\end{equation}
and the equation
\begin{equation}
\rv=1+\O
\end{equation}
holds globally [1]. Here $m(v)$ and $q^2(v)$ are arbitrary data
functions. It is explained in Ref. [1] why the regions of validity
of the two solutions are different.

The two metrics can be joined together on any line $L$ in the
region of overlap. For $L$ one can take a line of constant
$\rr$ with the value of $\rr$ small but satisfying (8). The
equation of this line in the coordinates $u,v$ can be
calculated, and it is
\begin{equation}
\displaystyle L\,\colon\qquad u=u_L(v)\;,\qquad
\frac{du_L(v)}{dv}=1+\O\;.
\end{equation}
To locate the line $L$, one then only needs to specify any point
through which it passes. Such a point is known [1]:
\begin{equation}
u=u_0\;,\qquad v=\Cv=v_0+4M\ln\frac{M^2}{\mu^2}+O(M)
\end{equation}
where $(u_0,v_0)$ is the point at which the apparent horizon
$\rr=0$ is null [1]. The junction conditions are
\begin{equation}
m(v)={\cal M}\left(u_L(v)\right)\;,\qquad
q^2(v)=Q^2\left(u_L(v)\right)\;.
\end{equation}

The initial values of the data functions are
\begin{equation}
m(v_0)=M\;,\qquad q^2(v_0)=0\;.
\end{equation}
The $\Cv$ in Eq. (20) is the bound set for $v$ by the correspondence
principle [1]. At $v\le\Cv$, the classical geometry is valid, and
the functions $m(v)$ and $q^2(v)$ differ from their values
in (22) microscopically.

The data functions are expressed through the metric in a
self-consistent manner. The metric should be transformed into the
null coordinates: $r=r(u,v)$, and the following function
should be calculated:
\begin{equation}
\kappa(u)=-\frac{d}{du}\ln\left.\left(-2\,\dr\right)\right|_{v=v_0}\;.
\end{equation}
The argument of $\ln$ in this expression is the red-shift factor.
The equations for the data functions as obtained in Ref. [1] are
of the form
\begin{equation}
-\frac{d{\cal M}(u)}{du}=\lambda\kappa^2(u)\;,\qquad
\frac{dQ^2(u)}{du}=2\lambda\kappa(u)\;,
\end{equation}
\begin{equation}
\lambda=\frac{\mu^2}{48\pi}\;.
\end{equation}
The numerical coefficient in (25) is calculated for the vacuum
of the massless spin-0 particles, and in (24) only the
contribution of the quantum s-mode is retained. The equation
for ${\cal M}(u)$ has also a contribution of higher-$l$ modes [1]
but it is by an order of magnitude smaller.

By derivation [1], Eqs. (24) are valid in the range of $u$ in
which two conditions are fulfilled:
\begin{equation}
\int\limits_{-\infty}^u du\,\kappa\gg 1\;,
\end{equation}
\begin{equation}
\frac{d}{du}\,\frac{1}{\kappa}\ll 1\;.
\end{equation}
Eq. (26) is a condition that the red shift is large. Eq. (27)
holds when $\kappa$ is a macroscopic quantity, and 
$d\kappa/du$ is a microscopic quantity. Both conditions are
fulfilled indeed during the entire radiation stage of the
evolution. At the post-radiation stage, condition (26) remains
valid but condition (27) does not. The point is that
$d\kappa/du$ remains a microscopic quantity but $\kappa\to 0$.
As $\kappa^2$ reaches the value of $d\kappa/du$, the quantity
in (27) becomes $O(1)$. Removal of condition (27) is another
problem that needs to be solved in the present work.

Condition (27) can be removed but the data equations get modified.
The refined equations are of the form
\begin{equation}
-\frac{d{\cal M}(u)}{du}=\lambda\left(\kappa^2(u)+
2\,\frac{d\kappa(u)}{du}-\frac{d\wkappa(u)}{du}\right)\;,
\end{equation}
\begin{equation}
\frac{dQ^2(u)}{du}=2\lambda\wkappa(u)
\end{equation}
where the function $\wkappa(u)$ is expressed through $\kappa(u)$
by the differential equation
\begin{equation}
\frac{d\wkappa}{du}=\wkappa\left(\wkappa-\kappa\right)
\end{equation}
with the boundary condition
\begin{equation}
\frac{1}{\wkappa(u)}\exp\left(-\int\limits_{-\infty}^u du\,
\kappa\right)\to 0\quad\;\mbox{as}\quad 
\int\limits_{-\infty}^u du\,\kappa\to\infty\;.
\end{equation}
Eq. (26) remains the only condition of validity of these
equations. In the range of $u$ in which $\kappa(u)$ satisfies
also condition (27), the solution of Eqs. (30), (31) is a series
\begin{equation}
\frac{1}{\wkappa}=\frac{1}{\kappa}\left(1+
\frac{d}{du}\,\frac{1}{\kappa}+
\frac{d}{du}\,\frac{1}{\kappa}\,\frac{d}{du}\,\frac{1}{\kappa}+
\cdots\right)
\end{equation}
in which all terms with derivatives are negligible. Then
$\wkappa=\kappa$, and one recovers Eqs. (24). The modification of 
the data equations is the most important fact for the presently
considered problem. Therefore, its derivation is sketched in
the Appendix below. The principal consequence of the modified
equations is that $\wkappa(u)$ cannot turn into zero at
a finite value of $u$.

The equations above make a closed system, and I go over to their
solution. Consider an outgoing light ray $\u$ that in the region
of strong field does not meet with microscopic $B$, and calculate
the minimum of $\rr$ on this ray. The equation of the minimum is [1]
\begin{equation}
B\rr=4A\;.
\end{equation}
Since $A=-|\O|$ and, on the ray considered, $B>|\O|$, one finds
that $\rr$ at the minimum is $\O$ and is negative. Therefore,
$\rr$ is $\O$ in the entire subregion of strong field foliated
by such light rays. It then follows from (15) that, in this subregion,
\begin{equation}
r=\rHplus (v)+\O\;,
\end{equation}
i.e., $r$ is a function only of $v$. Here and below, $\AHpm$ are
the two solutions for the apparent horizon given by Eq. (15):
\begin{equation}
\rHpm (v)=m(v)\pm\sqrt{m^2(v)-q^2(v)}\;.
\end{equation}
Since $r$ is a function only of $v$, so are $B$ and $A$ in (16)
and (17). Then consider the solution of Eq. (5):
\begin{equation}
-\ln\left(-2\,\dr\right)=
\int\limits_v^\infty dv\left.\frac{B}{2r}\right|_{\us}=
\int\limits_{v_L(u)}^\infty dv\left.\frac{B}{2r}\right|_{\us}+
\int\limits_v^{v_L(u)} dv\left.\frac{B}{2r}\right|_{\us}\;.
\end{equation}
A use of (34) enables one to calculate the strong-field contribution
in this integral:
\begin{equation}
\int\limits_v^{v_L(u)} dv\left.\frac{B}{2r}\right|_{\us}=
\int\limits_v^{v_L(u)} dv\,\frac{\BHplus(v)}{2\rHplus(v)}\;,
\end{equation}
\begin{equation}
\frac{d}{du}
\int\limits_v^{v_L(u)} dv\left.\frac{B}{2r}\right|_{\us}=
\frac{\BHplus\left(v_L(u)\right)}{2\rHplus\left(v_L(u)\right)}\;.
\end{equation}
In this way one recovers the strong-field metric of Ref. [1].

The weak-field contribution in (36) is readily obtained with
the aid of Eq. (10):
\begin{equation}
\int\limits_{v_L(u)}^\infty dv\left.\frac{B}{2r}\right|_{\us}=
-\ln\rr\Bigl|_{v=v_L(u)}\;,
\end{equation}
\begin{equation}
\frac{d}{du}
\int\limits_{v_L(u)}^\infty dv\left.\frac{B}{2r}\right|_{\us}=
\frac{1}{2r}\left.\left(\frac{4A}{\rr}-D\rr\right)\right|_L=\O\;.
\end{equation}
Combining (38) and (40) gives $\kappa$. Up to $\O$,
\begin{eqnarray}
\kappa(u)&\!=\!&
\frac{\BHplus\left(v_L(u)\right)}{2\rHplus\left(v_L(u)\right)}=
\left.\frac{\sqrt{m^2(v)-q^2(v)}}{\left(m(v)+\sqrt{m^2(v)-q^2(v)}
\right)^2}\right|_{v=v_L(u)}\nonumber\\
&\!=\!&\frac{\sqrt{{\cal M}^2(u)-Q^2(u)}}{\left({\cal M}(u)+
\sqrt{{\cal M}^2(u)-Q^2(u)}\right)^2}\;.
\end{eqnarray}
With this expression, Eqs. (24) close, and one recovers completely
the results of Ref. [1].

However, one obtains also the limitations on the validity of
these results. The region of their validity is bounded by the
latest light ray $\u$ that, in the strong field, does not meet
with small $B$. It can be shown that this upper bound in $u$
is the same as the one put by condition (27):
\begin{equation}
\kappa(u)>O(\sqrt{\lambda})\;,\qquad
{\cal M}^2(u)-Q^2(u)>|O(\lambda)|\;.
\end{equation}
By the junction conditions, the respective subregion of strong field
is bounded also from above in $v$:
\begin{equation}
m^2(v)-q^2(v)>|O(\lambda)|\;.
\end{equation}
By exploiting Eq. (7) we have removed one of these limitations.
Namely, under condition (43), the metric in (15)--(18) continues
the strong-field solution along the lines $\v$ across $B=0$
and $\AHmin$ down to $r=\O$, i.e., to the limit of validity
of semiclassical theory. The significance of the line $B=0$
is that, on it, $\rr$ has a minimum along the incoming light
rays $\v$, which enables these rays to cross the apparent
horizon twice. The main features of the strong-field solution 
are shown in Fig. 1 in the coordinates of a falling observer.
The subregion where the metric of Ref. [1] is valid is shown
foliated with the light rays $\u$ In the extended metric,
Eq. (33) has three solutions for each $v$ shown with
broken lines. Every outgoing ray $\u$ inside the apparent
horizon crosses one of these lines.

The values of $u$ and $v$ corresponding to the bounds (42)
and (43) (call them $u_1$ and $v_1$) are obtained in Ref. [1].
Up to negligible corrections,
\begin{equation}
u_1-u_0=96\pi\frac{M^3}{\mu^2}\;,\qquad
v_1-v_0=96\pi\frac{M^3}{\mu^2}\;.
\end{equation}
There remains to be considered the region $u>u_1$, $\,v>v_1$
marked with "?" in Fig. 1. At $u>u_1$, $\,\kappa(u)=O(\sqrt{\lambda})$
and, therefore, 
\begin{equation}
-\frac{d{\cal M}(u)}{du}=O(\lambda^2)\;,\qquad
\frac{dQ^2(u)}{du}=O(\lambda^{3/2})\;.
\end{equation}
It is seen that the flux of ${\cal M}$ is already at the two-loop 
level and is negligible as compared to the flux of $Q^2$.
Therefore, at $u>u_1$, $\,{\cal M}(u)$ may be considered constant:
\begin{equation}
{\cal M}(u)={\cal M}_1=\mbox{const.}\;,\qquad u>u_1\;.
\end{equation}
The flux of $Q^2$ is, on the other hand, significant
notwithstanding the fact that
\begin{equation}
{\cal M}_1{}^2-Q^2(u)=O(\lambda)\;,\qquad u>u_1\;.
\end{equation}
The value of ${\cal M}_1$ is obtained in Ref. [1].
${\cal M}_1$ is approximately 90\% of the ADM mass $M$.

In the strong-field sector of the region "?" one has accordingly
\begin{equation}
m(v)=m_1=\mbox{const.}\;,\quad m_1={\cal M}_1\;,\quad
m_1{}^2-q^2(v)=O(\lambda)\;,\quad v>v_1
\end{equation}
and, in addition,
\begin{equation}
B=\O\;,\qquad \rr=\O\;,\qquad r=m_1+\O\;.
\end{equation}
The last three conditions are equivalent in consequence of (48).
As a result, the metric in (15)--(17) simplifies as follows:
\begin{equation}
\rr=\frac{1}{4}B^2+\frac{q^2(v)-m_1{}^2}{m_1{}^2}\;,
\end{equation}
\begin{equation}
B=2\,\frac{r-m_1}{m_1}\;,\qquad
A=-\frac{1}{2m_1}\,\frac{dq^2(v)}{dv}\;.
\end{equation}
In order to transform this metric into the null coordinates,
differentiate Eq. (5) with $B$ inserted from (16), and next
use (48) and (49):
\begin{equation}
\frac{\partial^2}{\partial u\partial v}\ln\left(-\dr\right)=
\left(3\frac{q^2(v)}{r^4}-2\frac{m(v)}{r^3}\right)\dr=
\left(\frac{1}{m_1{}^2}+\O\right)\dr\;.
\end{equation}
In view of Eq. (18), this relation can also be written in the form
\begin{equation}
\frac{\partial^2}{\partial u\partial v}\ln\left(-
\frac{1}{\uv}\right)=
\left(\frac{1}{m_1{}^2}+\O\right)\frac{1}{\uv}\;,
\end{equation}
and it means that, in the strong-field sector of the region "?",
the Lorentzian subspace is the space of constant curvature.
Eq. (52) solves as
\begin{equation}
\dr=-2m_1{}^2\,\frac{f'(v)g'(u)}{\left(f(v)-g(u)\right)^2}\;,
\end{equation}
\begin{equation}
\frac{r-m_1}{m_1}=\frac{1}{2}B=m_1\,\frac{f''(v)}{f'(v)}
-2m_1\,\frac{f'(v)}{f(v)-g(u)}
\end{equation}
with the functions $f(v)$ and $g(u)$ to be fixed by the choice
of the coordinates $v$ and $u$. The primes on these functions
designate the differentiation with respect to their arguments.

The condition that fixes the normalization of $u$ is
\begin{equation}
\ru\Bigl|_L\equiv\rr\left.\left(2\,\dr\right)^{-1}\right|_L=-1\;.
\end{equation}
In a neighbourhood of the line $L$, $\,\rr$ should grow to
reach its weak-field value (8). This suggests that the line
$L$ is
\begin{equation}
\displaystyle L\,\colon\qquad f(v)=g(u)\;.
\end{equation}
With (57), condition (56) gets satisfied and, thereby, fixes
the function $g(u)$ as
\begin{equation}
g(u)=f\left(v_L(u)\right)\;.
\end{equation}
The condition that fixes the normalization of $v$ is (18), i.e.,
\begin{equation}
\frac{\partial r(u,v)}{\partial v}=\frac{1}{2}\,\rr\;.
\end{equation}
This gives the equation for $f(v)$
\begin{equation}
2\left(\frac{f''}{f'}\right)'-\left(\frac{f''}{f'}\right)^2
+\frac{m_1{}^2-q^2(v)}{m_1{}^4}=0\;.
\end{equation}
The equivalent equation for $g(u)$ is
\begin{equation}
2\left(\frac{g''}{g'}\right)'-\left(\frac{g''}{g'}\right)^2
+\frac{m_1{}^2-Q^2(u)}{m_1{}^4}=0\;.
\end{equation}
Both the metric in (54), (55) and the equations for $f$ and $g$
are invariant under the linear transformations
\begin{equation}
f\to \alpha f+\beta\;,\quad g\to \alpha g+\beta\;,\qquad
\alpha,\beta=\mbox{const.}
\end{equation}
and the transformation
\begin{equation}
f\to\frac{1}{f}\;,\quad g\to\frac{1}{g}\;.
\end{equation}
These transformations make a gauge arbitrariness of the solution for
$f$ and $g$.

For the red-shift factor in (23) one can write
\begin{equation}
\left.-2\,\dr\right|_{v=v_0}=\left.-2\,\dr\right|_{v=v_1}
\exp\left(-\int\limits_{v_0}^{v_1}dv\left.\frac{B}{2r}\right|_{\us}
\right)\;.
\end{equation}
To the integral in this expression, Eq. (34) applies. Therefore,
the exponential does not depend on $u$. When using (54) for the
remaining factor, it is convenient to make the replacement
\begin{equation}
\frac{1}{f(v_1)-g(u)}\to g(u)\;,
\end{equation}
thereby fixing the gauge with respect to the transformation (63).
As a result, $\kappa(u)$ is expressed through $g(u)$ in this
gauge as
\begin{equation}
\kappa(u)=-\frac{g''(u)}{g'(u)}\;.
\end{equation}
With this expression, Eq. (61) becomes the equation for $\kappa(u)$
\begin{equation}
-2\,\frac{d\kappa(u)}{du}-\kappa^2(u)+\frac{m_1{}^2-Q^2(u)}{m_1{}^4}
=0\;.
\end{equation}

Eqs. (29), (30), and (67) close. The closed system boils down to
two coupled equations for $\kappa$ and $\wkappa$ as functions
of $Q^2$. Upon the scaling
\begin{equation}
\kappa=\sqrt{\frac{2\lambda}{m_1{}^4}}\,k\;,\quad
\wkappa=\sqrt{\frac{2\lambda}{m_1{}^4}}\,\wk\;,\quad
\frac{Q^2-m_1{}^2}{2\lambda}=\delta\;,
\end{equation}
these equations take the form
\begin{equation}
-2\,\frac{dk}{d\delta}=\frac{k^2+\delta}{\wk}\;,\qquad
\frac{d\wk}{d\delta}=\wk-k\;.
\end{equation}
The function $Q^2(u)$ is expressed through their solution
by the exact integral
\begin{equation}
\mbox{const.}+u=\sqrt{\frac{m_1{}^4}{2\lambda}}\,\,
\frac{k^2-2k\wk+\delta}{\wk}\;,
\end{equation}
and for the functions $g(u)$ and $f(v)$ one obtains the exact 
expressions
\begin{equation}
g(u)=\exp\left(\frac{m_1{}^2-Q^2(u)}{2\lambda}\right)\;,\qquad
f(v)=\exp\left(\frac{m_1{}^2-q^2(v)}{2\lambda}\right)
\end{equation}
in which the gauge arbitrariness is already completely fixed.

One boundary condition to Eqs. (69) is (31), and it implies
\begin{equation}
\frac{1}{\wk}={\rm e}^\Delta\int\limits^\infty_\Delta
d\Delta\,{\rm e}^{-\Delta}\,\frac{1}{k}\;,\qquad
\Delta=\int\limits_{-\infty}^u du\,\kappa\;,\qquad
\frac{d\Delta}{d\delta}=\frac{k}{\wk}\;.
\end{equation}
The other one is that the solution should conform to (41):
\begin{equation}
k=\wk=\sqrt{-\delta}\;,\qquad \delta\to -\infty\;.
\end{equation}
It implies
\begin{equation}
\frac{dk}{d\delta}=-\frac{1}{2}\,{\rm e}^{-\textstyle\delta}
\int\limits_{-\infty}^{\textstyle\delta} d\delta\,
{\rm e}^{\textstyle\delta}\,\frac{1}{\wk}\;.
\end{equation}
From (74) one has $dk/d\delta<0$. Then from (72) one has
$\wk<k$, and then from (69) one has $d\wk/d\delta<0$.
This is sufficient to specify the solution.

The solution extends to $u=+\infty$. The Bondi charge $Q^2(u)$
grows monotonically up to the value
\begin{equation}
Q^2(u)=m_1{}^2+2\lambda\delta_0\;,\qquad 
\delta_0\le 0\;,\quad u\to+\infty
\end{equation}
and the function $\wkappa(u)$ decreases monotonically down to zero
according to the law
\begin{equation}
\wkappa(u)=\sqrt{\frac{2\lambda(-\delta_0)}{m_1{}^4}}
\exp\left(-\sqrt{\frac{2\lambda(-\delta_0)}{m_1{}^4}}\,\,u\right)\;,
\qquad \delta_0<0\;,\quad u\to+\infty\;.
\end{equation}
The function $\kappa(u)$ varies within the limits
\begin{equation}
\frac{\sqrt{m_1{}^2-Q^2(u)}}{m_1{}^2}<\kappa(u)<
\frac{\sqrt{m_1{}^2+2\lambda-Q^2(u)}}{m_1{}^2}
\end{equation}
and, at $u\to+\infty$, decreases down to a finite value:
\begin{equation}
\kappa(u)=\sqrt{\frac{2\lambda(-\delta_0)}{m_1{}^4}}+
\frac{\lambda}{m_1{}^4}\,u
\exp\left(-\sqrt{\frac{2\lambda(-\delta_0)}{m_1{}^4}}\,\,u\right)\;,
\qquad \delta_0<0\;,\quad u\to+\infty\;.
\end{equation}
For $\delta_0=0$, the laws of decrease are
\begin{equation}
\wkappa(u)=\frac{m_1{}^4}{\lambda}\,\frac{3}{u^3}\;,\qquad
\kappa(u)=\frac{3}{u}\;,\qquad\;\; \delta_0=0\;,\quad u\to+\infty\;.
\end{equation}
The difference between the cases $\delta_0<0$ and $\delta_0=0$
is unessential. The important point is that, by virtue of
the solution, $\delta_0$ is {\it necessarily nonpositive}.
The $-\delta_0$ is a pure number whose specific value is immaterial.

To summarize, the Bondi charges ${\cal M}^2(u)$ and $Q^2(u)$
begin with the values
\begin{equation}
{\cal M}^2(u)=M^2+\O\;,\quad\: Q^2(u)=\O\qquad\mbox{at}\quad u\le u_0
\end{equation}
and next draw together. They reach the level
\begin{equation}
{\cal M}^2(u)-Q^2(u)=|O(\lambda)|
\end{equation}
for a finite time $u$ in Eq. (44), but then they stay at this
level for an infinitely long time $u$, their fluxes
gradually decreasing down to zero.

Upon the use of (66) and (71), the expressions (54), (55) for
the strong-field metric take the form
\begin{equation}
\dr=-2m_1{}^2\wkappa(u)\wkappa\left(u_L(v)\right)
\frac{f(v)g(u)}{\left(f(v)-g(u)\right)^2}\;,
\end{equation}
\begin{equation}
r-m_1=-m_1{}^2\kappa\left(u_L(v)\right)+
2m_1{}^2\wkappa\left(u_L(v)\right)
\frac{f(v)}{f(v)-g(u)}\;.
\end{equation}
The data function $q^2(v)$ and hence the functions $f(v)$,
$\,\kappa\left(u_L(v)\right)$, $\,\wkappa\left(u_L(v)\right)$
are now known for all $v$ up to $v=\infty$. Specifically, the
behaviours of these functions at $v\to\infty$ are obtained
from (75)--(79) by replacing $u$ with $v$, and $Q^2(u)$
with $q^2(v)$.

Consider the limit $u\to+\infty$ along the rays $\v$ In this
limit, $g(u)={\rm e}^{-{\textstyle\delta}_0}$, and $r$ in (83)
is a finite function of $v$, while $\partial r/\partial u$
in (82) turns into zero. It follows that the null line
\begin{equation}
\mbox{EH}\,\colon\qquad g(u)={\rm e}^{-{\textstyle\delta}_0}
\end{equation}
is the event horizon. {\it The collapse finishes with a true
black hole.} The $\AHplus$ continues up to $v=\infty$ and,
in this limit, approaches the event horizon according to the law
\begin{equation}
\AHplus\,\colon\quad\left\{
\begin{array}{rcll}
g(u)-{\rm e}^{-{\textstyle\delta}_0}&\!\!=\!\!&\displaystyle
\exp\left(-2\,\sqrt{\frac{2\lambda(-\delta_0)}{m_1{}^4}}\,\,v\right)\;,
\qquad &\delta_0<0\;,\quad v\to\infty\\
g(u)-{\rm e}^{-{\textstyle\delta}_0}&\!\!=\!\!&\displaystyle
\sqrt{3}(2-\sqrt{3})\,\frac{m_1{}^4}{2\lambda}\,\frac{1}{v^2}\;,
\qquad &\delta_0=0\;,\quad v\to\infty\\
\end{array}
\right.
\end{equation}
or, equivalently,
\begin{equation}
\AHplus\,\colon\quad\left\{
\begin{array}{rcll}
v&\!\!=\!\!&\frac{1}{2}\,u\;,
\qquad &\delta_0<0\;,\quad u\to+\infty\\
v&\!\!=\!\!&\sqrt{p}\,\,u\;,\quad\displaystyle p=\frac{2-\sqrt{3}}{\sqrt{3}}\;,
\qquad &\delta_0=0\;,\quad u\to+\infty\;.\\
\end{array}
\right.
\end{equation}
Along both the EH and $\AHplus$,
\begin{equation}
\mbox{EH, }\AHplus\,\colon\qquad
r\to m_1+\sqrt{2\lambda(-\delta_0)}\;,\qquad v\to\infty\;.
\end{equation}
All outgoing light rays $\u$ with
$g(u)>{\rm e}^{-{\textstyle\delta}_0}$ cross the $\AHplus$
and go out to the future null infinity ($\I$).

In terms of the retarded time $g(u)$, the metric extends beyond
the event horizon and can be studied. The main facts are these.
The $\AHmin$ also continues up to $v=\infty$ and is
asymptotically of the form
\begin{equation}
\AHmin\,\colon\quad\left\{
\begin{array}{rcll}
{\rm e}^{-{\textstyle\delta}_0}-g(u)&\!\!=\!\!&\displaystyle
4{\rm e}^{-{\textstyle\delta}_0}
\sqrt{\frac{m_1{}^4(-\delta_0)}{2\lambda}}\,\frac{1}{v}\;,
\qquad &\delta_0<0\;,\quad v\to\infty\\
{\rm e}^{-{\textstyle\delta}_0}-g(u)&\!\!=\!\!&\displaystyle
\sqrt{3}(2+\sqrt{3})\,\frac{m_1{}^4}{2\lambda}\,\frac{1}{v^2}\;,
\qquad &\delta_0=0\;,\quad v\to\infty\\
\end{array}
\right.
\end{equation}
but, along it,
\begin{equation}
\AHmin\,\colon\qquad
r\to m_1-\sqrt{2\lambda(-\delta_0)}\;,\qquad v\to\infty\;.
\end{equation}
All outgoing light rays $\u$ with
$g(u)<{\rm e}^{-{\textstyle\delta}_0}$ cross the $\AHmin$ and
reach $v=\infty$. At $v=\infty$, all of them have the same
finite value of $r$:
\begin{equation}
g(u)=\mbox{const.}<{\rm e}^{-{\textstyle\delta}_0}\,\colon\qquad
r\to m_1-\sqrt{2\lambda(-\delta_0)}\;,\qquad\rr\to 0\;,\qquad v\to\infty\;.
\end{equation}
Their end-points at $v=\infty$ make a line which will be called
${\cal C}$. All of the said light rays cross ${\cal C}$ but,
for the metric to be obtained beyond ${\cal C}$, the data functions
$m(v)$ and $q^2(v)$ need to be known for "$v>\infty$". There is
nowhere to take these data from. ${\cal C}$ is a Cauchy horizon.

The chart $u$ ends at the event horizon. The chart $g(u)$ continues
farther but it has an end as well. It extends from the line $L$
on which $g(u)=f(v)$ down to $g(u)=-\infty$. In the limit
$g(u)\to -\infty$ along the rays $\v$, $\,r$ in (83) is a finite
function of $v$, while $\partial r/\partial g$ in (54) turns
into zero:
\begin{equation}
\frac{\partial r(u,v)}{\partial g(u)}=-2m_1{}^2\,
\frac{f'(v)}{\left(f(v)-g(u)\right)^2}\to 0\;,
\qquad g(u)\to -\infty\;.
\end{equation}
The null line $g(u)=-\infty$ will be called ${\cal R}$.
One can prove that this line is the light signal coming from the
singularity. It puts an end to the present consideration.
The rays $\v$ cross ${\cal R}$ but what is beyond ${\cal R}$
is also beyond the validity of semiclassical theory.

The full Penrose diagram is given in Fig. 2. The apparent horizon
is almost closed in a sense that almost all outgoing light rays
and almost all incoming ones cross it twice. The exceptions are
the event horizon and the rays $\v$ in the classical interval
of $v$ ($v\le\Cv$). The lines ${\cal R}$ and ${\cal C}$ bound
the region of validity of the present solution.

The boundary ${\cal R}$ is caused by the lack of knowledge
about the region of large curvature and does not present
an unsolvable problem. It is only the limit of validity
of semiclassical theory. If one has or will have a theory
valid at small scale, then it will remove this boundary.
For example, if one trusts the local terms of the
effective action even despite the fact that their
coefficients are uncertain, then the local polarization can be
calculated~[2]. It removes the singularity from the line
$r=0$ and shifts the apparent horizon away from this line.
The apparent horizon then closes in the sector between
the points~1 and~2 in Fig. 2.

The real problem is the unpredictability caused by the
Cauchy horizon ${\cal C}$ because this boundary is in
the region of validity of semiclassical theory and
cannot be removed. It is important to emphasize the
difference from the case of the Reissner--Nordstrom metric
whose analytic continuation is readily obtained. In that
case one deals with a {\it given} metric. When the metric
is obtained by integrating the evolution equations from
the initial state, as in the case of collapse, the Cauchy
horizon is an insurmountable barrier. In a sense, it is
a horizon for theoretical physics. The outgoing light rays
pass through it to another universe, and there is no way
to learn what will be with them next.
\newpage
\appendix
\setcounter{equation}{0}

{\renewcommand{\theequation}{A.\arabic{equation}}

\begin{center}
\subsection*{\bf Appendix. Derivation of the modified data equations.}  
\end{center}

$$ $$

It suffices to consider the contributions of the s-mode (${\bf\Psi}_0$)
to the fluxes of ${\cal M}$ and $Q^2$ because only the s-mode
contributes to the flux of $Q^2$ [1], and only the flux of $Q^2$ is
significant when $\kappa=\O$. For these contributions one has [1]
\begin{eqnarray}
-\partial_u {\cal M}&\!\!=\!\!&
\langle\,(\partial_u {\bf\Psi}_0)^2\rangle\biggl|_{\I}
-\xi\,\partial^2_{uu}\langle\,{\bf\Psi}_0{}^2\rangle\biggl|_{\I}\;,\\
\partial_u Q^2&\!\!=\!\!&
(1-4\xi)\,\partial_u\langle\,{\bf\Psi}_0{}^2\rangle\biggl|_{\I}\;.
\end{eqnarray}
Here $\xi$ is the parameter of the scalar-field equation [1] 
which in 4 dimensions is $1/6$, and in 2 dimensions is zero.
Both expectation values in (A.1), (A.2) are expressed through
the same spectral function [1]:
\begin{equation}
\langle\,(\partial_u {\bf\Psi}_0)^2\rangle\biggl|_{\I}
=\frac{2}{(4\pi)^2}\int\limits_0^\infty d\eout\,I_0(\eout,u)
+\mbox{c.c.}\;,
\end{equation}
\begin{equation}
\langle\,{\bf\Psi}_0{}^2\rangle\biggl|_{\I}
=\frac{2}{(4\pi)^2}\int\limits_0^\infty \frac{d\eout}{{\rm i}\eout}
\int\limits_{-\infty}^u d{\bar u}\,I_0(\eout,{\bar u})
\exp\left({\rm i}\eout(u-{\bar u})\right)+\mbox{c.c.}\;,
\end{equation} 
\begin{equation}
I_0(\eout,u)=\int\limits_0^\infty d\ein\,{\dot U}(u)
\int\limits_{-\infty}^{u+0} du'\,\ein{\dot U}(u')
\exp\Bigl({\rm i}(\Omega-\Omega')\Bigr)\;,
\end{equation}
\begin{equation}
\Omega-\Omega'=\ein\left(U(u)-U(u')\right)+\eout(u-u')\;,
\end{equation}
\begin{equation}
{\dot U}(u)\equiv\frac{dU(u)}{du}=
\exp\left(-\int\limits_{-\infty}^u du\,\kappa\right)\;.
\end{equation}
The calculation needs to be done under condition (26).

In (A.5) make the replacement of the integration variables
\begin{equation}
y=\ein{\dot U}(u)\frac{1}{\wkappa(u)}\;,\qquad
x=\ein{\dot U}(u')\frac{1}{\wkappa(u')}\;.
\end{equation}
For the needed Jacobian to emerge, $\wkappa(u)$ should satisfy
Eq. (30), and, in order that the lower limit in $x$ could be
set equal to zero, $\wkappa(u)$ should satisfy the boundary
condition (31). With the function $\wkappa(u)$ thus defined,
one obtains
\begin{equation}
I_0(\eout,u)=\wkappa(u)\left[1+P\left(u,{\rm i}z,\frac{d}{d{\rm i}z}
\right)\right]F(z)
\end{equation}
where
\begin{equation}
z=\frac{\eout}{\wkappa(u)}\;,\qquad
F(z)=\frac{2\pi z{\rm e}^{\textstyle -\pi z}}{
{\rm e}^{\textstyle \pi z}-{\rm e}^{\textstyle -\pi z}}\;,
\end{equation}
and the function $P$ is defined as follows. The equation
\begin{equation}
\ln\frac{x}{y}=\int\limits^u_{u'}du''\,\wkappa(u'')
\end{equation}
following from (A.8) and (30) should be solved with respect
to the quantity
\begin{equation}
\wkappa(u)(u-u')=\ln\frac{x}{y}+f\Bigl(u,\ln\frac{x}{y}\Bigr)\;.
\end{equation}
The function $P$ is expressed through $f$ in (A.12) as
\begin{eqnarray}
P\left(u,{\rm i}z,\ln\frac{x}{y}\right)&\!\!=\!\!&
\exp\left({\rm i}zf\Bigl(u,\ln\frac{x}{y}\Bigr)\right)-1\nonumber\\
&\!\!=\!\!&{}
\sum_{k,p}h_{k,p}(u)({\rm i}z)^k\left(\ln\frac{x}{y}\right)^{k+p}\;,
\qquad k\ge 1\;,\quad p\ge 1
\end{eqnarray}
and is a series of the form (A.13).

Only the real part of $I_0(\eout,u)$, and, therefore, only the
even $p$ in the series (A.13) contribute to the expectation
value (A.3). Upon the insertion of (A.9) and (A.13) with
$p=2n$ in the spectral integral (A.3), this integral boils
down to
\begin{equation}
\int\limits_0^\infty dz\,({\rm i}z)^k\left(\frac{d}{d{\rm i}z}
\right)^{k+2n}F(z)={\rm i}(-1)^k k!\left(
\frac{d}{d{\rm i}z}\right)^{2n-1}F(z)\Biggr|_{\textstyle z=0}\;.
\end{equation}
The even powers of $z$ in $F(z)$ drop out of this expression.
Only the odd part of $F(z)$ contributes, and this odd part is
\begin{equation}
\frac{1}{2}\left(F(z)-F(-z)\right)=-\pi z\;.
\end{equation}
It follows that only the terms with $p=2$ in the series (A.13)
contribute to the expectation value (A.3), and this contribution
can be calculated:
\begin{equation}
\langle\,(\partial_u {\bf\Psi}_0)^2\rangle\biggl|_{\I}
=\frac{1}{48\pi}\left(\wkappa^2-\frac{2}{\wkappa}\,
\frac{d^2\wkappa}{du^2}+\frac{3}{\wkappa^2}\,
\left(\frac{d\wkappa}{du}\right)^2\right)\;.
\end{equation}
Eq. (30) for $\wkappa$ can now be used to obtain finally
\begin{equation}
\langle\,(\partial_u {\bf\Psi}_0)^2\rangle\biggl|_{\I}
=\frac{1}{48\pi}\left(\kappa^2+2\,\frac{d\kappa}{du}\right)\;.
\end{equation}
If this expression is inserted in (A.1) with $\xi=0$, the
result for $\partial_u{\cal M}$ will be precisely the one
that the 2-dimensional effective action gives. This is
a good check.

For the calculation of the expectation value (A.4), introduce
in (A.4) the new integration variables
\begin{equation}
\gamma=\eout(u-{\bar u})\;,\qquad
\sigma=\int\limits^u_{\bar u} du''\,\wkappa(u'')\;,
\end{equation}
and, as discussed in Ref. [1], set $\wkappa(u)=\kappa(u)=0$
for $u<u_0$. Then (A.4) will take the form
\begin{equation}
\langle\,{\bf\Psi}_0{}^2\rangle\biggl|_{\I}
=\frac{2}{(4\pi)^2}\int\limits_0^\infty
\frac{d\gamma}{{\rm i}\gamma}\,{\rm e}^{\textstyle {\rm i}\gamma}
\int\limits_0^\Gamma d\sigma\left(1+P\left({\bar u},{\rm i}{\bar z},
\frac{d}{d{\rm i}{\bar z}}\right)\right)F({\bar z})+\mbox{c.c.}
\end{equation}
where
\begin{equation}
\Gamma=\int\limits^u_{u_0} du\,\wkappa\;,\qquad
{\bar z}=\frac{\gamma}{\wkappa({\bar u})(u-{\bar u})}\;,
\end{equation}
and the denominator of ${\bar z}$ should be expressed through 
$\sigma$. This expression (with the obvious change of notation)
is just (A.12). It is only important that
\begin{equation}
{\bar z}=O\left(\frac{\gamma}{\sigma}\right)\to 0
\quad\;\mbox{as}\quad\sigma\to\infty
\end{equation}
since one only needs the asymptotics of (A.19) at 
$\Gamma\to\infty$. From the leading asymptotics, the contribution
of $P$ drops out entirely, and one obtains
\begin{equation}
\langle\,{\bf\Psi}_0{}^2\rangle\biggl|_{\I}
=\frac{4}{(4\pi)^2}\left(\int\limits_0^\infty d\gamma\,
\frac{\sin\gamma}{\gamma}\right)F(0)\,\Gamma
=\frac{1}{8\pi}\Gamma\;,\qquad\Gamma\to\infty
\end{equation}
\begin{equation}
\partial_u\langle\,{\bf\Psi}_0{}^2\rangle\biggl|_{\I}
=\frac{1}{8\pi}\,\wkappa(u)\;,\qquad
\int\limits^u_{u_0} du\,\wkappa\to\infty\;.
\end{equation}
That $\Gamma\to\infty$ is the needed limit, i.e., that
$\Gamma\to\infty$ follows from (26) can be seen from
the integrated Eq. (30):
\begin{equation}
\exp\left(-\int\limits^u_{u_0} du\,\wkappa\right)
=\frac{\wkappa(u_0)}{\wkappa(u)}
\exp\left(-\int\limits^u_{u_0} du\,\kappa\right)\;.
\end{equation}
Since $\wkappa(u_0)=O(1)$, this quantity vanishes in the
limit (26) by virtue of the boundary condition (31).
Eqs. (A.1), (A.2) with $\xi=1/6$ and the expectation values
inserted from (A.17) and (A.23) are the equations presented
in the text.
}

\newpage
\begin{center}
\section*{\bf References}
\end{center}

$$ $$

\begin{enumerate}
\item G.A. Vilkovisky, Ann. Phys. 321 (2006) 2717 [hep-th/0511182];
Phys. Lett. B 634 (2006) 456 [hep-th/0511183]; Phys. Lett. B 638
(2006) 523 [hep-th/0511184]. 
\item V.P. Frolov and G.A. Vilkovisky, Phys. Lett. B 106 (1981)
307; in {\it Proc. 2nd Seminar on Quantum Gravity, Moscow, 1981}
(M.A. Markov and P.C. West, Eds.), p. 267, Plenum, London, 1983. 
\end{enumerate}

\newpage
\begin{center}
\section*{\bf Figure captions}
\end{center}

$$ $$

\begin{itemize}
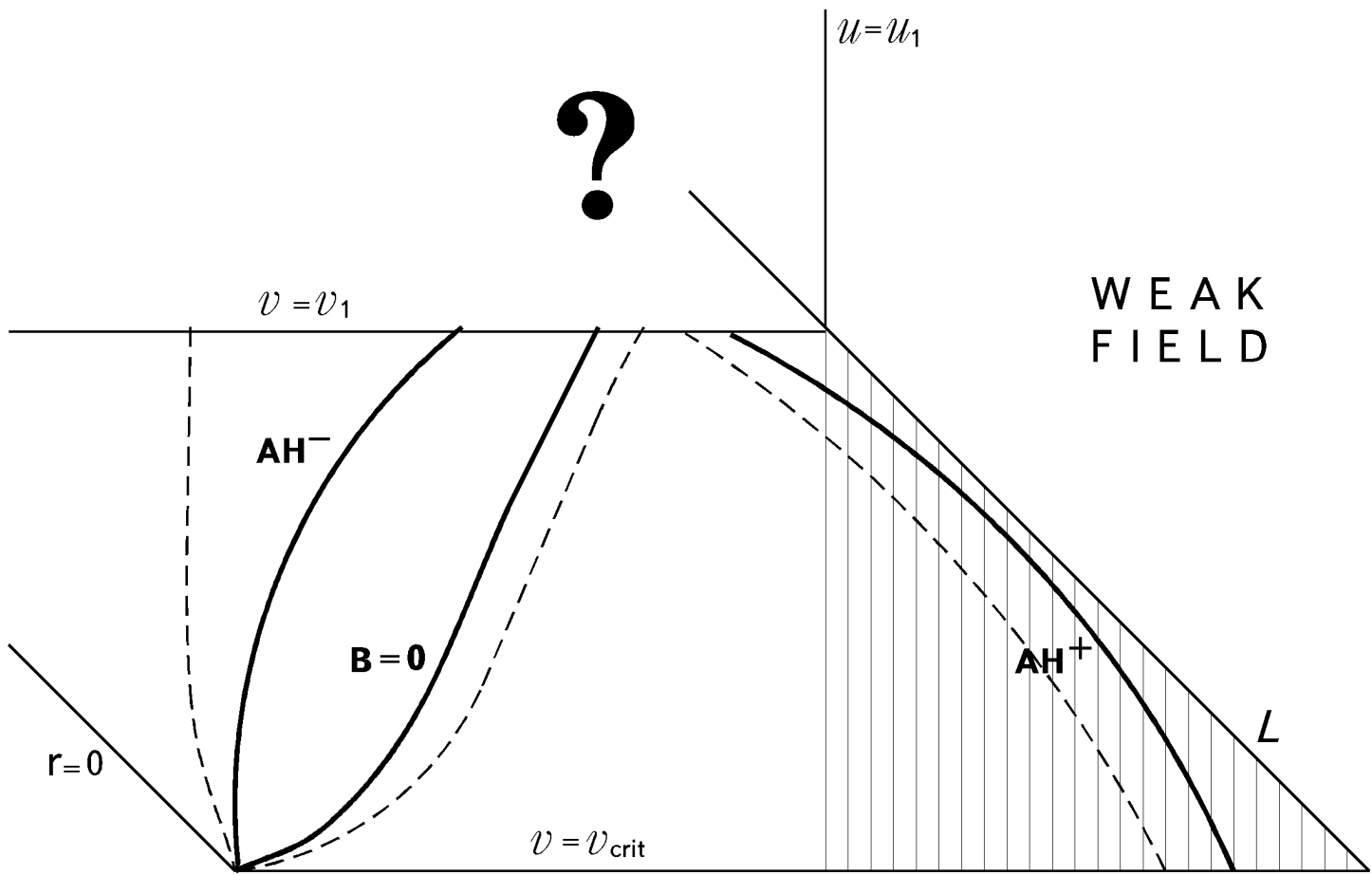
\item[Fig.1.] The region of strong field restricted to
condition (43). The horizontal lines are $\v$ The vertical
lines are $\u$ The three bold curves are $\AHmin$, $\,B=0$,
and $\AHplus$. The three broken curves make a line of extrema
(minima) of $\rr$ along the rays $\u$ $\,L$ is the border
between the strong-field and weak-field regions.
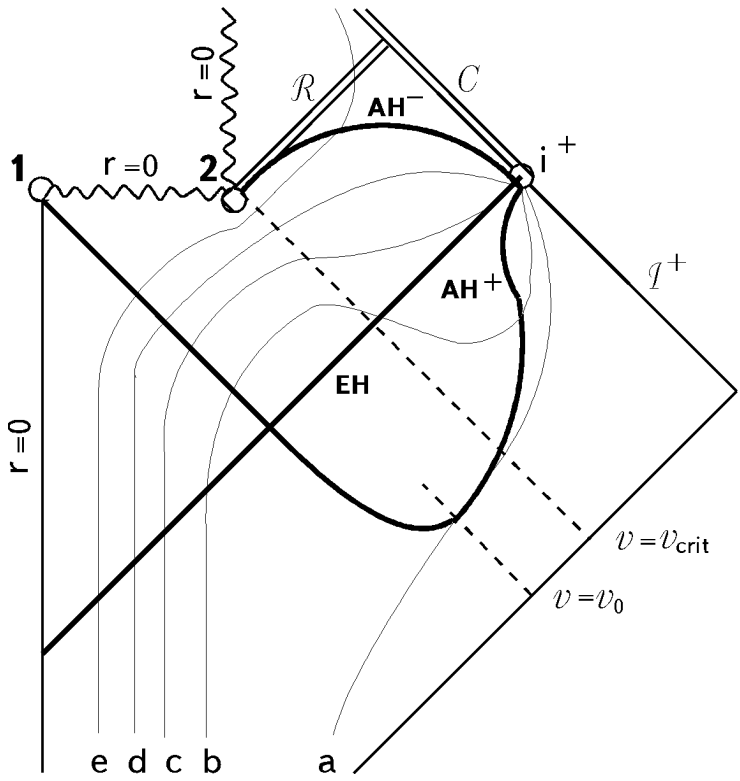
\item[Fig.2.] Penrose diagram for the collapse spacetime.
The bold lines are $\AHmin$, EH, and $\AHplus$. The double
lines are the boundaries ${\cal R}$ and ${\cal C}$.
The wavy lines mean the singularity. The light curves are
the level lines of $r$: (a)~$r=2M$, (b)~$r=2M-|\O|$,
(c)~$r=m_1+\sqrt{2\lambda(-\delta_0)}$,
(d)~$r=m_1-\sqrt{2\lambda(-\delta_0)}$,
(e)~$r<m_1-\sqrt{2\lambda(-\delta_0)}$.
\end{itemize}

\end{document}